\documentstyle[aps,manuscript]{revtex}

\begin{document}

\title{Growth and magnetic properties of MnO$_{2-\delta}$ nanowire microspheres}

\author{J.B. Yang, X.D. Zhou, W.J. James}
\address{Graduate Center for Materials Research and Departments
of Physics and Chemistry, University of Missouri-Rolla, Rolla, MO
65409}
\author{S.K. Malik}
\address{Tata Institute of Fundamental Research, Colaba, Mumbai 400-005, India}
\author{C. S. Wang}
\address{School of Physics, Peking University, Beijing 100871, P.R. China}

\maketitle
\begin{abstract}

We report the synthesis of MnO$_{2-\delta}$ microspheres using hydrothermal and 
conventional chemical reaction methods. The microspheres of MnO$_{2-\delta}$ consist of 
nanowires having a diameter of 20-50 nm and a length of 2-8 $\mu$ m. The value of the 
oxygen vacancy $\delta$ estimated from the x-ray photoelectron spectrum (XPS) is ~0.3. 
The magnetization versus temperature curve indicates a magnetic transition at about 13 
K. It is found that a parasitic ferromagnetic component is imposed on the 
antiferromagnetic structure of MnO$_{2-\delta}$, which might result from distortion of 
the lattice structure due to oxygen vacancies. The magnetic transition temperature 
T$_N$ is about 10 K lower than that of the bulk MnO$_2$ single crystal.

\end{abstract}

\newpage

The possibility of controlling the structure and the chemical composition of materials 
at the nanoscale level is of great interest for both basic science and technological 
applications. In particular, one dimensional(1-D) nanometric structures of oxides have 
attracted intensive attention\cite{daun,hu,heath,yang}. For example MnO$_2$ has 
distinctive properties which have enabled its use as catalysts, ion-sieves, and 
electrode materials\cite{ram,amm,hill}. Both $\alpha$- and $\gamma$- MnO$_2$ are 
poential candidates for cathodes and catalysts in Li/MnO$_2$ batteries \cite{ram}. They 
can be converted into Li$_{1-x}$Mn$_2$O$_4$ cathode by electrochemical Li$^+$ 
intercalation, in which Li$^+$ was inserted or extracted during the charging or 
discharge process. MnO$_2$ nanowires are of great interest due to that their morphology 
simultaneously minimize the distance over which Li$^+$ must diffuse during the 
discharge and charging processes\cite{li}. This may provide the opportunity to 
determine the theoretical operating limits of a Lithium battery as the 1-D nanowire is 
the smallest structure for the efficient electron transport. MnO$_2$ nanowires or 
nanostructure materials also contain much highly surface area which may become more 
ideal host materials for the insertion and extraction of lithium ions and chemical 
reactions. Therefore, well defined 1-D nanostructures appear as a much better candidate 
for studies on the space-confined transport phenomena as well as applications. 
Considerable effort has been made to prepare bulky or nanocrystalline MnO$_2$ with 
different structures\cite{bach,hunter,ros,wu}. Recently, single crystal MnO$_2$ 
nanowires have been synthesized using a hydrothermal method\cite{wang1}. MnO$_2$ also 
possesses an interesting magnetic structure\cite{zhuang,kaw,sato,yama}. Up to now, very 
few investigations on the magnetic properties of nano-sized Mn-oxides have been 
reported. In this letter, we report the synthesis of microspheres of MnO$_{2-\delta}$ 
consisting of nanowires. The microspheres were prepared using both hydrothermal and 
room temperature chemical reaction methods. The structure, morphology, composition and 
magnetic properties of the Mn-oxide microspheres were studied using x-ray 
diffraction(XRD), scanning electron microscopy(SEM), transmission electron 
microscopy(TEM), magnetic measurements and x-ray photoelectron spectroscopy(XPS). 

~~~The materials synthesized here were prepared by oxidation of hydrated manganese 
sulfate MnSO$_4$.H$_2$O with an equal amount of ammonium persulfate 
(NH$_4$)$_2$S$_2$O$_8$\cite{wang1}. MnSO$_4$.H$_2$O (0.08mol) and 
(NH$_4$)$_2$S$_2$O$_8$ (0.08 mol) were dissolved in 150 ml distilled water at room 
temperature to form a clear solution. Half of the solution was transferred into a 
stainless steel autoclave, sealed and maintained at 120 $^\circ$ for 12 h. Another half 
of the solution was kept at room temperature in air for 6 h. After the reaction, the 
resulting black solid from each solution was filtered and washed with distilled water 
and acetone several times, then dried at 60 $^\circ$C for 24 h. X-ray diffraction (XRD) 
using Cu-$K_{\alpha}$ radiation shows the material to be single phase, 
$\alpha$-MnO$_2$. The magnetization curves of the samples were measured using a SQUID 
magnetometer in a field of up to 6 T from 1.5 K to 300 K. The morphology was studied 
using SEM(JEOL-6340F) and TEM. X-ray photoelectron spectra were collected using a 
"KRATOS" model AXIS 165 XPS spectrometer with a Mg source and an Al monochrometer.

~~~Fig. 1 shows the x-ray diffraction patterns of the samples obtained by both methods. 
The diffraction peaks can be indexed to $\alpha$-MnO$_2$ with lattice parameters 
a=9.784$\AA$ and c=2.863$\AA$, space group I4/m. The XRD pattern of the room 
temperature treated sample(Fig. 1(b)) shows much broader, and less intensive peaks, due 
to its smaller grain size and the many crystal defects. There are some changes in the 
relative peak intensities in Fig. 1(b), suggesting a preferred orientation among the 
Mn-oxide grains. The mean size of the particles (Fig. 1(a)) is about 40 nm as 
determined by the Scherrer formula using the width of the [211] peak from Fig. 1 (a), 
and was further confirmed by SEM and TEM imaging(Fig. 2). 

Fig. 2 shows the SEM((a)-(d)) and TEM(e) images of the samples prepared by the two 
methods. Figs. 2(a) and (b) show a sea urchin-like sphere with a diameter of 4-8 $\mu$m  
for the sample prepared by the hydrothermal method. These microspheres consist of 
bundles of small wires with a diameter of 20-50 nm(see TEM image (e)). Figures 2(c) and 
(d) show the typical morphology of the MnO$_2$ obtained by the reaction carried out at 
room temperature. The diameter of these microspheres varies from 1 to 4 $\mu$m. The 
microspheres are also made up of smaller needle-like wires, however, they are not so 
sharply defined as compared to the samples made by the hydrothermal method. The 
hydrothermal treatment increases the diameter of the microspheres and the length of the 
nanowires. It is found that the concentration of the solution is very important in 
forming the microspheres. The microspheres can be synthesized when the Mn$^{2+}$ 
concentration is higher than 0.3 mol/L. It suggests that, at high Mn$^{2+}$ 
concentrations, the formation of microspheres by aggregation of the nanowires is 
favored by a decrease in the total surface energy of the system.
      
A typical room temperature x-ray photoelectron spectrum of the Mn$_{2-\delta}$ prepared 
by the hydrothermal method is plotted in Fig. 3. The peaks of Mn (3s, 3p, 2p, 2s) and O 
(1s, 2s) are observed. Auger peaks from Mn and O are also observed as O KLL and Mn LMM 
lines. The binding energies of the Mn 2p3/2 and 2p1/2 states are 641.5 and 653.6 ev, 
respectively, which are lower than those of the  standard MnO$_2$(642.1 and 653.8 ev) 
\cite{xps}. The decrease in binding energy is likely due to the oxygen vacancies formed 
in these materials which would decrease the Mn-O bond strength. The estimated relative 
ratio of Mn to O is 37:63 corresponding to a composition of about MnO$_{1.7}$.  
       
Fig. 4 shows the zero field cooling (ZFC) and field cooling(FC) magnetization curves of 
Mn$_{2-\delta}$ prepared by the hydrothermal method measured under different magnetic 
fields. A kink is observed at about 13 K in Fig. 4 (a) and (b), corresponding to the 
N\'eel magnetic  transition temperature T$_N$. This temperature is lower than that of a 
MnO$_2$ single crystal \cite{yama} T$_N$=24.5 K, which is due to the small grain size 
effect\cite{muly}. The kink as well as the difference between the ZFC and FC curves 
becomes smaller as the applied magnetic field increases. The inset in Fig. 4 (a) shows 
the ZFC and FC curves of the sample prepared at room temperature. A similar phenomenon 
is observed, except that the transition peaks are much broader and the magnetization 
value is higher which may be due to more defects and a larger distortion in this 
sample.  Fig. 5 is the magnetization versus applied magnetic field curves of the sample 
prepared by the hydrothermal method. At 1.8 K, the M-H curve shows a small hysteresis, 
which indicates that a ferromagnetic component is superposed on the antiferromagnetic 
curve. This agrees with the M-T curves in that the magnetization increases with 
decreasing temperature suggesting a ferrimagnetic or ferromagnetic structure is formed 
in this compound. At 300 K, nearly linear M-H curves reveal that a paramagnetic state 
exists in this temperature range. The ferromagnetic component at low temperature may 
result from the noncollinear molecular field that cants the two antiparallel 
sublattices due to the distortion of the crystal structure, especially  arising from 
the oxygen vacancies. The small canted angle between the moments of the Mn sublattices 
leads to a small ferromagnetic moment\cite{Nai}.

\section{Summary}

MnO$_{2-\delta}$ microspheres have been synthesized using hydrothermal and room 
temperature chemical reaction methods. The microspheres consist of MnO$_{2-\delta}$ 
nanowires having a diameter of 20-50 nm and a length of 2-8$\mu$m. The estimated value 
of the oxygen vacancy $\delta$ from XPS data is 0.3. The magnetic measurements indicate 
that MnO$_2$ has an antiferromagnetic structure with a parasitic ferromagnetic 
component below the N\'eel temperature, T$_N$.  T$_N$ is about 13 K which is 10 K lower 
than that of bulk MnO$_2$ single crystals.

\section{ACKNOWLEDGMENTS}

The support by DOE under DOE contract \#DE-FC26-99FT400054 is
acknowledged. We would like to thank A.G.S. Hemantha and Prof. J. Sweitz for help with 
the magnetic measurements.

\begin{figure}
\caption{X-ray diffraction patterns of the sample prepared by the hydrothermal method 
(a) and room temperature reaction (b).}
\end{figure}

\begin{figure}
\caption{ The SEM (a-d) and TEM(e) images of the Mn-oxides: (a), (b) prepared by 
hydrothermal method; (c) and (d) prepared at room temperature; (e) TEM image prepared 
by hydrothermal method.}
\end{figure}

\begin{figure}
\caption { Room temperature x-ray photoelectron spectrum of MnO$_{2-\delta}$ prepared 
by the hydrothermal method. }
\end{figure}

\begin{figure}
\caption { The temperature dependence of the magnetization curves
under zero field cooling (ZFC) and field cooling (FC) for MnO$_{2-\delta}$ prepared by 
the hydrothermal method: (a) H=50 Oe, (b) H=1 kOe, (c) H=10 kOe). The inset in Fig. 
4(a) is for the sample made at room temperature
}
\end{figure}

\begin{figure}
\caption {The magnetization versus applied magnetic field curves of MnO$_{2-\delta}$  
prepared by the hydrothermal method at different temperatures. }
\end{figure}

\end{document}